\documentclass[traditabstract]{aa}

\usepackage{natbib}
\bibpunct{(}{)}{;}{a}{}{,} 
\usepackage{amssymb}
\usepackage{url}
\usepackage{graphicx}
\usepackage{longtable}
\usepackage[usenames,dvipsnames]{color}

\newcommand{\submm}{submillimetre}

\newcommand{\urltt}[1]{\texttt{#1}}
\newcommand{\micron}{\mbox{$\mu$m}}
\newcommand{\msun}{\mbox{$M_\odot$}}
\newcommand{\lsun}{\mbox{$L_\odot$}}
\newcommand{\msunyr}{\mbox{\msun\ yr$^{-1}$}}

\begin{document}

 \title{Spatially-resolved dust properties of the \\
GRB 980425 host galaxy\thanks{{\it Herschel} is an ESA space observatory with science instruments provided by European-led Principal Investigator consortia and with important participation from NASA.},\thanks{Reduced images as FITS files are only available at the CDS via anonymous ftp to \protect\url{cdsarc.u-strasbg.fr (130.79.128.5)} or via
\protect\url{http://cdsarc.u-strasbg.fr/viz-bin/qcat?J/A+A/562/A70}}
 }
 
\titlerunning{Spatially-resolved dust properties of the GRB 980425 host galaxy}
\authorrunning{Micha{\l}owski et al.}

\author{Micha{\l}~J.~Micha{\l}owski\inst{\ref{inst:gent},\ref{inst:roe}}\thanks{FWO Pegasus Marie Curie Fellow}
\and
L.~K.~Hunt\inst{\ref{inst:hunt}}
\and
E.~Palazzi\inst{\ref{inst:pal}}
\and
S.~Savaglio\inst{\ref{inst:sav}}
\and
G.~Gentile\inst{\ref{inst:gent},\ref{inst:brus}}
\and
J.~Rasmussen\inst{\ref{inst:dark},\ref{inst:dtu}}
\and
M.~Baes\inst{\ref{inst:gent}}
\and
S.~Basa\inst{\ref{inst:basa}}  
\and
S.~Bianchi\inst{\ref{inst:hunt}}
\and
S.~Berta\inst{\ref{inst:sav}}
\and
D.~Burlon\inst{\ref{inst:sydney}}
\and
J.~M.~Castro~Cer\'{o}n\inst{\ref{inst:jm}} 
\and
S.~Covino\inst{\ref{inst:cov}}
\and
J.-G.~Cuby\inst{\ref{inst:basa}}
\and
V.~D'Elia\inst{\ref{inst:elia1},\ref{inst:elia2}} 
\and
P.~Ferrero\inst{\ref{inst:fer1},\ref{inst:fer2}}
\and
D.~G\"otz\inst{\ref{inst:sacley}}
\and
J.~Hjorth\inst{\ref{inst:dark}}
\and
M.~P.~Koprowski\inst{\ref{inst:roe}}
\and
D.~Le Borgne\inst{\ref{inst:leb}} 
\and
E.~Le Floc'h\inst{\ref{inst:sacley}} 
\and
D.~Malesani\inst{\ref{inst:dark}}
\and
T.~Murphy\inst{\ref{inst:sydney}} 
\and
E.~Pian\inst{\ref{inst:pal}}
\and
S.~Piranomonte\inst{\ref{inst:pir}} 
\and
A.~Rossi\inst{\ref{inst:ros}}
\and 
J.~Sollerman\inst{\ref{inst:soll}}
\and
N.~R.~Tanvir\inst{\ref{inst:tan}}
\and
A.~de Ugarte Postigo\inst{\ref{inst:ant},\ref{inst:dark}}
\and
D.~Watson\inst{\ref{inst:dark}}
\and
P.~van der Werf\inst{\ref{inst:vdw}}
\and
S.~D.~Vergani\inst{\ref{inst:cov}}
\and
D.~Xu\inst{\ref{inst:dark}} 
	}

\institute{
Sterrenkundig Observatorium, Universiteit Gent, Krijgslaan 281-S9, 9000, Gent, Belgium  \label{inst:gent}\\
e-mail: {\tt mm@roe.ac.uk}
\and
SUPA\thanks{Scottish Universities Physics Alliance}, Institute for Astronomy, University of Edinburgh, Royal Observatory, Edinburgh, EH9 3HJ, UK
\label{inst:roe}
\and
INAF-Osservatorio Astrofisico di Arcetri, Largo E. Fermi 5, I-50125 Firenze, Italy \label{inst:hunt}
\and
INAF-IASF Bologna, Via Gobetti 101, I-40129 Bologna, Italy \label{inst:pal}
\and
Max-Planck-Institut f\"{u}r Extraterrestrische Physik, Giessenbachstra{\ss}e, D-85748 Garching bei M\"{u}nchen, Germany \label{inst:sav}
\and
Department of Physics and Astrophysics, Vrije Universiteit Brussel, Pleinlaan 2, 1050 Brussels, Belgium \label{inst:brus}
\and
Dark Cosmology Centre, Niels Bohr Institute, University of Copenhagen, Juliane Maries Vej 30, DK-2100 Copenhagen \O, Denmark  \label{inst:dark}
\and
Technical University of Denmark, Department of Physics, Frederiksborgvej 399, DK-4000 Roskilde, Denmark \label{inst:dtu}
\and
Aix Marseille Universit\'e, CNRS, LAM (Laboratoire d'Astrophysique de Marseille) UMR 7326, 13388, Marseille, France \label{inst:basa}
\and
Sydney Institute for Astronomy, School of Physics, The University of Sydney, NSW 2006, Australia \label{inst:sydney}
\and
Herschel Science Centre (ESA-ESAC), E-28.692 Villanueva de la Ca\~nada (Madrid), Spain \label{inst:jm}
\and
INAF/Osservatorio Astronomico di Brera, via Emilio Bianchi 46, 23807 Merate (LC), Italy \label{inst:cov}
\and
ASI Science Data Centre, Via Galileo Galilei, 00044 Frascati (RM), Italy \label{inst:elia1}
\and
INAF - Osservatorio Astronomico di Roma, Via di Frascati, 33, 00040 Monteporzio Catone, Italy \label{inst:elia2}
\and
 Instituto de Astrof\'{\i}sica de Canarias (IAC), E-38200 La Laguna, Tenerife, Spain \label{inst:fer1}
\and
   Departamento de Astrof\'{\i}sica, Universidad de La Laguna (ULL), E-38205 La Laguna, Tenerife, Spain \label{inst:fer2}
\and
Laboratoire AIM-Paris-Saclay, CEA/DSM/Irfu - CNRS - Universit\'e Paris Diderot, CE-Saclay, pt courrier 131, F-91191 Gif-sur-Yvette, France \label{inst:sacley}
\and
Institut d'Astrophysique de Paris, UMR 7095, CNRS, UPMC Univ. Paris 06, 98bis boulevard Arago, F-75014 Paris, France \label{inst:leb}
\and
INAF - Osservatorio Astronomico di Roma, via Frascati 33, 00040 Monte Porzio Catone (RM), Italy \label{inst:pir}
\and
Th\"uringer Landessternwarte Tautenburg, Sternwarte 5, D-07778 Tautenburg, Germany \label{inst:ros}
\and
The Oskar Klein Centre, Department of Astronomy, AlbaNova, Stockholm University, 106 91 Stockholm, Sweden \label{inst:soll} 
\and
Department of Physics and Astronomy, University of Leicester, University Road, Leicester, LE1 7RH, UK \label{inst:tan}
\and
Instituto de Astrof\' isica de Andaluc\' ia (IAA-CSIC), Glorieta de la Astronom\' ia s/n, E-18008, Granada, Spain \label{inst:ant}
\and
Leiden Observatory, Leiden University, P.O. Box 9513, NL-2300 RA Leiden, The Netherlands \label{inst:vdw}
}

\abstract{
Gamma-ray bursts (GRBs) have been proposed as a tool for studying star formation in the Universe, so it is crucial to investigate whether their host galaxies and immediate environments are in any way special compared with other star-forming galaxies. Here we present spatially resolved maps of dust emission of the host galaxy of the closest known GRB 980425 at $z=0.0085$ using our new high-resolution observations from {\it Herschel}, Atacama Pathfinder Experiment (APEX), Atacama Large Millimeter Array (ALMA) and Australia Telescope Compact Array (ATCA). 
We modelled the spectral energy distributions of the host and of the star-forming region displaying the Wolf-Rayet signatures in the spectrum (WR region), located $800$ pc  from the GRB position. 
{ The host is characterised by low dust content and a high fraction of UV-visible star formation, similar to other dwarf galaxies. These galaxies are abundant in the local universe, so it is not surprising to find a GRB in one of them, assuming the correspondence between the GRB rate and star formation.}
The WR region contributes substantially to the host emission  at the far-infrared, millimetre, and radio wavelengths and we propose that this is a consequence of its high gas density. If dense environments are also found close to the positions of other GRBs, then the ISM density should also be considered, along with metallicity, an important factor influencing whether a given stellar population can produce a GRB. 
}


\maketitle

\section{Introduction}
\label{sec:intro}

{
Long ($\mbox{duration}>2$~s) gamma-ray bursts (GRBs) have been shown to be associated with the death of massive stars (e.g.~\citealt{hjorthnature,stanek}; for a review see \citealt{hjorthsn}), and therefore signal very recent star formation in their host galaxies. 
{ This is also supported by the detection of the signatures of Wolf-Rayet (WR) stars in GRB hosts { (\citealt{hammer06,han10}, but see \citealt{chen07,schulze11}),  by very low ages of the GRB sites \citep{christensen08,thone08,ostlin08}, and by the concentration of GRBs in the UV-brightest pixels of their hosts \citep{bloom02,fruchter06,kelly08}, similar to that of WR stars and supernovae Ic \citep{leloudas10}. Theoretically, rotating WR stars are the best candidates for long GRB progenitors, mainly because of their masses, angular momentum, and lack of the hydrogen envelope \citep{woosley,woosley06,yoon05,yoon06}. }}
However, the debate about whether the GRB rate is proportional to the star formation rate (SFR) density in the Universe is still ongoing \citep{conselice05,castroceron06,fruchter06,stanek06,wainwright07, yuksel08,fynbo09,kistler09,savaglio09,leloudas10,leloudas11,levesque10c,levesque10,svensson10,elliott12,hjorth12,michalowski12grb,robertson12,boissier13,hao13,perley13}. 

{ 
Since the presence of dust is intimately connected to star formation, we would expect to find
measurable amounts of dust in GRB hosts.}
The characteristics of the dusty medium can be studied from its emission at long wavelengths, in particular in the far-infrared (far-IR) and {\submm} regimes that trace the emission of colder ($<60$ K) dust, which constitutes most of the total dust mass of a galaxy. However, because of insufficient sensitivity of previous instruments, only four GRB hosts have been detected at {\submm} wavelengths  (\citealt{frail}, \citealt{berger}, \citealt{tanvir}, \citealt{wang12}; see the full compilation of {\submm} observations in \citealt{deugartepostigo12}).

It is only now, with the successful operation of the {\it Herschel}
Space Observatory 
\citep{herschel}
 and the Atacama Large Millimeter Array (ALMA), that the dust in a significant number of GRB hosts can be 
studied, as these facilities are capable of not only detecting its emission, but also mapping its spatial distribution.}
{ This is crucial in order to understand the interplay between the star formation giving rise to the GRB progenitor and the dust formation and to understand the physical conditions necessary for a massive star to explode as a GRB. 
 It has been claimed that GRBs reside in low-metallicity environments, which could be reflected in the distribution of dust 
within their hosts.}

Mapping the spatial distribution of dust
is limited to the closest and most extended GRB hosts, so an obvious candidate for 
a resolved dust study is \object{GRB 980425} at redshift $z=0.0085$ \citep{tinney98} and its associated supernova \object{SN 1998bw} \citep{galamanature} located in the barred spiral galaxy \object{ESO 184-G82}. As this galaxy hosted the closest known GRB, it has been the subject of numerous studies \citep{fynbo00, sollerman05,foley06,hammer06,stanek06,hatsukade07,lefloch,lefloch12,christensen08,michalowski09,michalowski12grb,savaglio09,castroceron10}.
It  is a dwarf ($0.02 L_B^*$) barred spiral   showing a large number of star-forming regions \citep{fynbo00,sollerman05};
SN\,1998bw occurred inside one of these \citep{fynbo00,sollerman02}, $\sim800$ pc southeast of a region displaying a Wolf-Rayet type spectrum  \citep[hereafter: WR region]{hammer06}. This region dominates the galaxy's emission at $24\,\micron$ \citep{lefloch}, contributes substantially to its SFR \citep[but not to its stellar mass;][]{michalowski09},  is {young { \citep[$1$--$6$ Myr;][]{hammer06,christensen08}}, and  exhibits the lowest metallicity among star-forming regions  within the host with $ 12 + \log(\mbox{O}/\mbox{H})=8.16$, i.e. $0.3$ solar \citep{christensen08} compared with $12 + \log(\mbox{O}/\mbox{H})=8.6$, i.e. $0.8$ solar \citep{sollerman05} for the entire host { \citep[a common characteristic in nearby GRBs; Fig.~3 of][]{levesque11,thone08}}. { Highly ionised mid-IR emission lines of the WR region also suggest a very young episode of star formation \citep{lefloch12}.}

{Because GRB 980425 ended its life in a region with very few massive stars which does not exhibit any WR spectral signatures,  \citet{hammer06} have proposed that the GRB progenitor had been expelled from the nearby WR region with a velocity of the order of $\sim100$--$300\,\mbox{km s}^{-1}$, not atypical of Galactic runaway stars \citep{blaauw93,tenjes01,dewit05}.
}

Even for this nearby and well studied GRB host the cold dust properties are not very well constrained, because the currently available infrared observations of this galaxy are limited to wavelengths shortward of $160\,\mu$m \citep{lefloch12}.
The objective of this paper is to change this situation by determining the dust properties 
in this closest known GRB host.
This is possible with our new high-resolution data ranging from far-IR to radio.
}

This paper is structured as follows.
In Sect.~\ref{sec:data} we describe our data. 
Then, in Sect.~\ref{sec:sed}, we describe the spectral energy distribution (SED) modelling we apply to these data, and present our results in Sect.~\ref{sec:res}.
We discuss the implications of our results in 
Sect.~\ref{sec:discussion}, and close with a summary of our results in Sect.~\ref{sec:conclusion}.
We use a cosmological model with $H_0=70$ km s$^{-1}$ Mpc$^{-1}$,  $\Omega_\Lambda=0.7$, and $\Omega_m=0.3$, so GRB 980425 at $z=0.0085$ is at a luminosity distance of 36.5 Mpc and $1''$ corresponds to 175 pc at its redshift (i.e.~our resolution of $\sim7''$ at $100\,\micron$ corresponds to 1.3 kpc).

\section{Data}
\label{sec:data}

We obtained {\it Herschel} observations of the GRB 980425 host (project no.~OT1\_lhunt\_2, PI: L.~Hunt) using the Photodetector Array Camera and Spectrometer  \citep[PACS;][]{pacs} and the Spectral and Photometric Imaging Receiver \citep[SPIRE;][]{spire} 
with a total integration time of 1116 and 445 s, respectively, on 13--14 Mar 2011\footnote{OBSIDs: 1342216005, 1342216055 and 1342216056}. 
The {\it Herschel} observations of the entire GRB host sample within this program are presented in \citet{hunt14}. 

We used PACS in Small-Scan map mode (20\arcsec/s),
with ten scan legs, 3\arcmin\ long, separated by 4\arcsec\ steps.
The scans were divided into two  Astronomical Observation Requests (AORs), 
with orthogonal scan directions that were executed sequentially. 
With this configuration we obtained homogeneous coverage over an area with 
a diameter of $\sim2$\arcmin, sufficient to cover the region subtended
by the GRB\,980425 host galaxy.
Cross scans gave the needed 
redundancy to 
avoid 1/f noise and spurious detector glitches 
on 
science and noise maps.
Our estimated 1$\sigma$ sensitivity is 0.9 and 2.1\,mJy at 100 and 160\,$\mu$m, respectively.
With SPIRE in Small-Map Mode, we used three repetitions in order to obtain
a sensitivity of roughly the 1$\sigma$ confusion limit
\citep[see][]{nguyen10} of $\sim$6\,mJy beam$^{-1}$ at 250\,$\mu$m.

Data reduction for PACS and SPIRE was performed with {\sc Hipe} \citep[Herschel Imaging Processing Environment;][]{hipe}
v10.0. 
For PACS, we used {\sc Hipe} to obtain the initial calibration and reduction, and
then {\it scanamorphos} \citep[v16.0;][]{scanamorphos} 
to make the final maps,
in order to better account for faint, extended emission.
We used pixel sizes of 1\farcs7 and 3\farcs00 for PACS 100 and 160\,$\mu$m,
and 4\farcs5, 6\farcs25, and 9\farcs0 for SPIRE 250, 350, and 500\,$\mu$m, respectively. 

In each PACS and SPIRE image, an estimate for the background was obtained by averaging the flux measured within a set of sky apertures close to the galaxy but far enough away to avoid the contamination from the galaxy emission.

{
After background subtraction, the flux densities of the entire host galaxy at all {\it Herschel} wavelengths were obtained in an aperture of $50$\arcsec\  diameter. }
{
The aperture corrections were negligible in all cases when compared with calibration uncertainties \citep{ciesla12,dale12}.

Because of limitations in image resolution it was possible to measure reliably the WR region flux only for the PACS data at 100 $\mu$m and 160 $\mu$m where the source is point-like. An aperture with a radius of $6$\arcsec\   
was used to measure the flux and a standard PSF extrapolation was applied for aperture correction.
}

The uncertainties associated with the measured photometric values reported in Table~\ref{tab:data} were computed as a combination in quadrature of the calibration uncertainty, 7\% for SPIRE data (according to Version 2.4, 7 June 2011, of the SPIRE Observer's Manual) and 5\% for PACS data (according to Version 2.3, 8 June 2011  of the PACS Observer's Manual), and the sky uncertainty derived by considering the number of pixels within the galaxy and the WR region apertures and the standard deviation of the average value in the individual sky apertures \citep[see][]{dale12}.
{
Before addition in quadrature, the sky uncertainties were corrected for the correlated noise, since the pixels are not independent.}

We also performed submillimetre ($870\,\mu$m) observations of the GRB 980425 host on 4--5 Aug 2007 (project no.~O-079.F-9312A-2007, PI: M.~Micha{\l}owski) using the Large Apex BOlometer CAmera \citep[LABOCA;][]{laboca} mounted at the Atacama Pathfinder Experiment \citep[APEX;][]{apex}. A total of 4.1 hr of on-source data were obtained. During most of the observations the weather was excellent with $0.25$ mm of precipitable water vapour. Data reduction was done using the {\sc Crush}\footnote{\url{www.submm.caltech.edu/~sharc/crush/}} package \citep{crush}. We used the `deep'  option that results in the best signal-to-noise ratio for faint objects. We checked that the `faint' option, with weaker filtering, resulted in a higher noise level. The analysis was done using the {\sc Miriad} package \citep{miriad,miriad2}. The beam size for the final image was $19\farcs5$ and the rms was $1.4$ mJy beam$^{-1}$.  The target was detected and is extended at the resolution of APEX/LABOCA with the measured full width at half maximum (FWHM) of $(35\pm11)''\times(30\pm9)''$, and deconvolved size of $29''\times23''$, consistent with the size at optical wavelengths (half-light diameter at $R$-band of $22''$, i.e.~$4$ kpc). The flux for the entire galaxy was found by summing the signal over a $50$\arcsec\  diameter aperture.
The resolution of the data was not sufficient to separate the WR region from the rest of the galaxy.

{

We performed Band 3 ALMA observations on 1 Sep 2012 (project no.~2011.0.00046.S, PI: M.~Micha{\l}owski). A total of 67.4 min of on-source data were obtained. Four $1.875$ GHz spectral windows were centred at  $100.6$,  $102.4$, $112.5$, and $114.3$ GHz. 
Twenty three antennas and baselines ranging between $24$ and $384$ m were available. Neptune, J1733-130, and J1945-552 were used as flux, bandpass, and phase calibrators, respectively. The amount of precipitable water vapour ranged between $1.8$--$2.15$ mm. The data reduction and analysis were done using the {\sc Casa} package \citep{casa}. The 
map was created by multi-frequency synthesis with `natural' weighting, giving the best signal-to-noise ratio. 
The WR region is the only 
detected star-forming region in the GRB 980425 host. We measured its flux by fitting a Gaussian at its position, whereas the flux for the entire galaxy was 
again measured in the $50$\arcsec\  diameter aperture.

We performed radio 2.1 GHz observations with the Australia Telescope Compact Array (ATCA) on 12 Apr 2012  with a total integration time of 12 hr (project no.~C2700, PI: M.~Micha{\l}owski). The array was in the 1.5B configuration with baselines up to 1500 m. 
We used 2 GHz bandwidth.
The data reduction and analysis were done using {\sc Miriad}. 
We analysed the data separately in four $0.5$ GHz frequency ranges centred at 1324, 1836, 2348, and 2860 MHz. The WR region is detected at all  four of these frequencies dominating the radio flux of the GRB 980425 host. Again, we used a Gaussian fit and a $50$\arcsec\  diameter aperture to measure the flux of the WR region and the entire galaxy, respectively.

We also obtained Ultraviolet/Optical Telescope \citep[UVOT;][]{uvot} data from the  {\it Swift} \citep{swift} archive\footnote{\url{http://heasarc.gsfc.nasa.gov/cgi-bin/W3Browse/swift.pl}}. The GRB 980425 host was observed on 15-16 Jun 2007  (obsid 00036585001) and on 18 Jun 2007  (obsid 00036585002) with total integration times of $2034$, $1350$, and $1014$ s for the {\it uvw2} ($0.20\,\micron$), {\it uvm2} ($0.22\,\micron$), and {\it uvw1} ($0.26\,\micron$) filters, respectively. We only used these filters, as 
 the data at other (optical) filters are superseded by the ground-based data. We measured the fluxes of the entire galaxy in a $50$\arcsec\ aperture and derived the aperture corrections from the curve-of-growth analysis. For  the WR region we used a standard aperture with a radius of $5$\arcsec. We added in quadrature $0.03$ mag to the errors to account for zero-point uncertainty\footnote{\url{http://heasarc.gsfc.nasa.gov/docs/swift/analysis/uvot_digest/zeropts.html}.}.  We corrected the fluxes for Galactic extinction assuming $E(\mbox{B$-$V})=0.059$ mag \citep{schlegel98,sollerman05} and the extinction curve of \citet{cardelli89}.
The photometry is fully consistent with lower resolution GALEX data \citep{michalowski09}.

 Finally, we used the Wide-field Infrared Survey Explorer (WISE) All-Sky images\footnote{\url{http://wise.ssl.berkeley.edu/astronomers.html}.} \citep{wise}. Again, for the entire galaxy we used a $50$\arcsec\ aperture and the curve-of-growth analysis. For the WR region we used standard apertures (radius $8.25$\arcsec\ for band 1--3 and $16.25$\arcsec\ for band 4) and applied appropriate aperture corrections\footnote{\url{http://wise2.ipac.caltech.edu/docs/release/allsky/expsup/}}. Our fluxes are consistent with those from {\it Spitzer} \citep{lefloch} for overlapping filters.
 }

\begin{figure*}
\begin{center}
\includegraphics[width=\textwidth,viewport=0 403 344 820,clip]{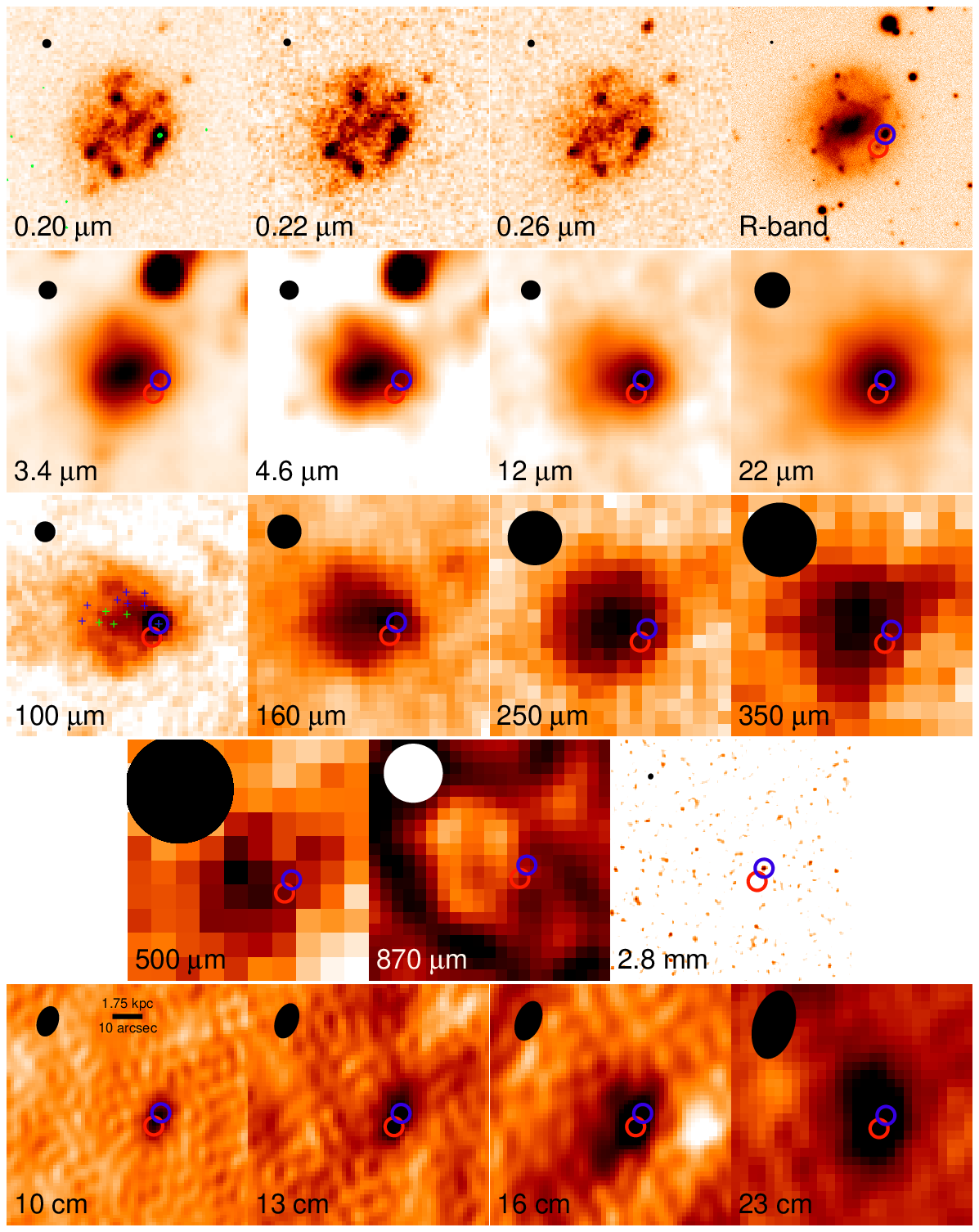}
\end{center}
\caption{Mosaic of the images of  the GRB 980425 host.
The $R$-band is from \citet{sollerman05} and the remaining images are presented here for the first time. North is up and east is to the left. Each panel is  $80''\times80''$  ($14\mbox{ kpc}\times14\mbox{ kpc}$). { The scale is indicated on the 10 cm image}, { whereas the FWHM beamsize is shown on each panel as a {\it filled circle or ellipse}.} The {\it red} and {\it blue  circles} 
show the GRB position and the WR region, respectively. 
{ {\it Blue} and {\it green crosses} on the $100\,\micron$ image show the star-forming regions with high electron density $n_e> 145\,\mbox{cm}^{-3}$ and  $\mbox{age}\ge6$ Myr, respectively (see Sect.~\ref{sec:wr}).} 
{ The ALMA $2.8$~mm image is shown as $3.5$, $4$, and $5\sigma$ contours on the $0.2\,\mu$m image.}
The $870\,\micron$ image has been inverted.
}
\label{fig:im}
\end{figure*}

The UVOT, WISE, {\it Herschel}, LABOCA, ALMA, and ATCA images are shown in Fig.~\ref{fig:im} and the results of the photometry are summarised in Table~\ref{tab:data}. { A shift in the flux centroid to the west from $3.4\,\micron$ to $22\,\micron$ and to the east from $160\,\micron$ to $500\,\micron$ is a result of the increasing \citep{lefloch} and decreasing contribution of the WR region to the total flux, respectively.}

For the SED modelling we also used the compilation of the data presented in \citet[][their Table~1]{michalowski09} covering ultraviolet (UV) to radio wavelengths \citep[including the data from][]{watson04,sollerman05, lefloch,castroceron10}. We also used the $70\,\micron$ photometry of the entire host from \citet{lefloch12}, but not 
at $160\,\micron$, as these data are superseded by our higher-resolution and more sensitive data. { Their measurements are consistent with ours within the uncertainties.}

\begin{table}
\scriptsize
\caption{Our new UVOT, WISE, {\it Herschel} / PACS and SPIRE, APEX / LABOCA, ALMA and ATCA photometry results, the contribution of the WR region and the beam sizes. The limit is $2\sigma$. \label{tab:data}}
\begin{center}
\begin{tabular}{l r@{$\pm$}l r@{$\pm$}l c c l}
\hline\hline
$\lambda_{\rm obs}$ & \multicolumn{2}{c}{Entire host} & \multicolumn{2}{c}{WR region} & \% & FWHM & Inst. \\
(\micron) & \multicolumn{2}{c}{(mJy)} & \multicolumn{2}{c}{(mJy)} & & ($''$)\\
\hline
0.2033 & $1.70$ & $0.05$ &  $0.15$ & $0.01$ & 9 & 2.92 & UVOT \\   
0.2229 & $1.70$ & $0.05$ &  $0.15$ & $0.01$ & 9 & 2.45 & UVOT \\   
0.2591 & $1.79$ & $0.06$ &  $0.14$ & $0.01$ & 8 &  2.37 & UVOT \\   
3.4 & $4.3$ & $0.3$ &  $0.54$ & $0.04$ & 1.3   & 6.1 & WISE \\
4.6 & $2.3$ & $0.3$ &  $0.32$ & $0.03$ & 14  &  6.4 & WISE\\
12 & $11.9$ & $0.8$ &  $2.1$ & $0.1$ & 18  &  6.5 & WISE\\
22 & $28.7$ & $0.9$ &  $16$ & $1$ & 56  &  12 & WISE\\
100 & $327$ & $12$ &  $72$ & $5$ & 19 & 6.7$\times$6.9 & PACS\\
160  & $323$ & $19$ & $74$ & $9$ & 13 & 10.7$\times$12.1 & PACS\\
250 & $231$ & $16$ & \multicolumn{2}{c}{$\cdots$} & $\cdots$ & 18.2 & SPIRE\\
350 & $106$ & $8$ & \multicolumn{2}{c}{$\cdots$}& $\cdots$ & 24.9 & SPIRE\\
500 & $43$ & $4$& \multicolumn{2}{c}{$\cdots$}& $\cdots$ & 36.3  & SPIRE\\
870 & $12.7$ & $1.8$& \multicolumn{2}{c}{$\cdots$}& $\cdots$ & 19.5 & APEX\\
2790.5 & \multicolumn{2}{c}{$<0.43$} & $0.123$ & $0.025$ & $>29$ & 1.9$\times$1.8 & ALMA\\ 
2790.5 & \multicolumn{2}{c}{$\cdots$} & $0.069^{\mathrm a}$ & $0.028$ & $\cdots$ & 1.9$\times$1.8 & ALMA\\
1.05e5  & $0.40$ & $0.20$ & $0.28$ & $0.03$ & 70 & 10.3$\times$7.0 & ATCA\\
1.28e5  &  $0.71$ & $0.19$ & $0.29$ & $0.03$ & 41 & 12.0$\times$7.5 & ATCA\\
1.63e5  & $1.11$ & $0.21$ & $0.42$ & $0.04$ & 38 & 13.5$\times$8.3 & ATCA\\
2.26e5  & $0.84$ & $0.16$ & $0.60$ & $0.06$ & 71 & 23$\times$14 & ATCA\\
\hline
\end{tabular}
$^{\mathrm a}$~With free-free emission subtracted as described in Sect.~\ref{sec:sed}.
\end{center}
\end{table}

\section{SED fitting}
\label{sec:sed}

{ Before fitting the SED, we estimated the free-free emission \citep{caplan86} in the WR region using
the H$\alpha$ flux measured by \citet{christensen08} corrected for Galactic extinction.
The estimated free-free flux at 2.8\,mm is 0.054\,mJy, $\sim$45\% of the total observed WR flux.
Hence, we used in the fit at 2.8\,mm 0.069\,mJy, which is the remainder after correcting
for the free-free emission.
The 2.8\,mm free-free emission over the entire galaxy is found to be $\sim$0.21\,mJy, well under
the upper limit of our non-detection.
At $870\,\micron$ we estimate a total free-free contribution of $\sim$0.19\,mJy, less than 2\% 
of the total observed flux and smaller than the calibration uncertainties; hence we neglect its correction
at this wavelength.
}

We applied the SED fitting method detailed in \citet[][see therein a discussion of the derivation of galaxy properties and typical uncertainties]{michalowski08,michalowski09,michalowski10smg,michalowski10smg4} based on 35\,000 templates in the library of \citet{iglesias07}, plus additional templates of \citet{silva98} and \citet{michalowski08}, all developed in {\sc Grasil}\footnote{\url{www.adlibitum.oat.ts.astro.it/silva}} \citep{silva98}. They are based on numerical calculations of radiative transfer within a galaxy that is assumed to be a triaxial system with diffuse dust and dense molecular clouds, in which stars are born.

The templates cover a broad range of galaxy properties from quiescent to starburst. 
Their star formation histories are assumed to represent a smooth Schmidt-type law \citep[SFR proportional to the gas mass to some power; see][for details]{silva98} with a starburst (if any) in addition starting $50$ Myr before the time at which the SED is computed. There are seven free parameters in the library of \citet{iglesias07}: the normalization of the Schmidt-type law, the timescale of the mass infall, the intensity of the starburst, the timescale of the molecular cloud destruction, the optical depth of molecular clouds, the age of a galaxy, and the inclination of a disk with respect to the observer.

We also used {\sc Magphys}\footnote{\url{www.iap.fr/magphys}} \citep[Multi-wavelength Analysis of Galaxy Physical Properties;][]{dacunha08}, an empirical but physically-motived SED model, which is based on the energy balance between the energy absorbed by dust and re-emitted in the infrared. { We used the \citet{bruzualcharlot03} stellar models and adopted the \citet{chabrier03} initial mass function (IMF).} Similarly to {\sc Grasil}, in {\sc Magphys} two dust media are assumed: diffuse interstellar medium (ISM) and dense stellar birth clouds. Four dust components are considered: cold dust ($15$--$25$ K), warm dust ($30$-$60$ K), hot dust ($130$--$250$ K), and polycyclic aromatic hydrocarbons (PAHs).

\renewcommand{\tabcolsep}{0.1cm}
\begin{table*}
\caption{{\sc Magphys} results from the SED fitting. \label{tab:magphysres}}
\scriptsize
\begin{center}
\begin{tabular}{lcccccccccccccc}
\hline\hline
Region & $\log L_{\rm IR}$ & SFR & sSFR & $\log M_*$ & $\log M_d$ & $\tau_V$ & $T_{\rm cold}$ & $\xi_{\rm cold}$ & $T_{\rm warm}$ & $\xi_{\rm warm}$ & $\xi_{\rm hot}$ & $\xi_{\rm PAH}$ &  $f_\mu$ \\
 & ($L_\odot$) & ($M_\odot\,\mbox{yr}^{-1}$) & (Gyr$^{-1}$) & ($M_\odot$) & ($M_\odot$) & & (K) & & (K) & & & & \\
(1) & (2) & (3) & (4) & (5) & (6) & (7) & (8) & (9) & (10) & (11) & (12) & (13) & (14) \\
\hline
Entire host & $9.012^{+0.005}_{-0.005}$ & $0.26^{+0.08}_{-0.08}$ & $0.53$ & $8.68^{+0.30}_{-0.30}$ & $6.21^{+0.02}_{-0.01}$ & $0.16$ & $17.2^{+0.8}_{-0.0}$ & $0.22^{+0.05}_{-0.01}$ & $52^{+4}_{-1}$ & $0.53^{+0.04}_{-0.01}$ & $0.14^{+0.00}_{-0.03}$ & $0.082^{+0.012}_{-0.000}$ & $0.30^{+0.02}_{-0.02}$ &  \\
WR region & $8.407^{+0.015}_{-0.000}$ & $0.02^{+0.01}_{-0.01}$ & $2.37$ & $7.00^{+0.30}_{-0.30}$ & $4.93^{+0.29}_{-0.25}$ & $0.47$ & $21.7^{+1.5}_{-2.7}$ & $0.18^{+0.00}_{-0.05}$ & $60^{+1}_{-1}$ & $0.67^{+0.05}_{-0.01}$ & $0.09^{+0.00}_{-0.01}$ & $0.071^{+0.000}_{-0.005}$ & $0.19^{+0.01}_{-0.03}$ &  \\
\% & 25 & 7 & $\cdots$ &2 & 5 & $\cdots$ &$\cdots$ &$\cdots$ &$\cdots$ &$\cdots$ &$\cdots$ &$\cdots$ &$\cdots$ & \\
Host$-$WR & $8.772^{+0.005}_{-0.005}$ & $0.19^{+0.06}_{-0.06}$ & $0.27$ & $8.82^{+0.30}_{-0.30}$ & $5.84^{+0.25}_{-0.26}$ & $0.10$ & $18.9^{+2.6}_{-2.3}$ & $0.24^{+0.08}_{-0.07}$ & $35^{+5}_{-3}$ & $0.47^{+0.07}_{-0.08}$ & $0.16^{+0.02}_{-0.02}$ & $0.129^{+0.011}_{-0.010}$ & $0.36^{+0.05}_{-0.05}$ &  \\
\hline
\end{tabular}
\tablefoot{(1) the entire host galaxy / only the WR region / percentage contribution of the WR region to the cumulative properties of the galaxy / the host with WR fluxes subtracted. (2) $8-1000\,\mu$m infrared luminosity. (3) star formation rate from SED modelling. (4) specific star formation rate ($\equiv\mbox{SFR}/M_*$). (5) stellar mass. (6) dust mass. (7) average $V$-band optical depth ($A_V=1.086\tau_V$).  (8) temperature of the cold dust component. (9) contribution of the cold component to the infrared luminosity. (10) temperature of the warm dust component. (11) contribution of the warm component to the infrared luminosity. (12) contribution of the hot ($130$--$250$ K, mid-IR continuum) component to the infrared luminosity. (13) contribution of the PAH component to the infrared luminosity. (14) contribution of the ISM dust (as opposed to birth clouds) to the infrared luminosity.}
\end{center}
\end{table*}

\section{Results}
\label{sec:res}

\subsection{Basic physical properties}
\label{sec:prop}

\begin{figure*}
\begin{center}
\includegraphics[width=\textwidth,clip]{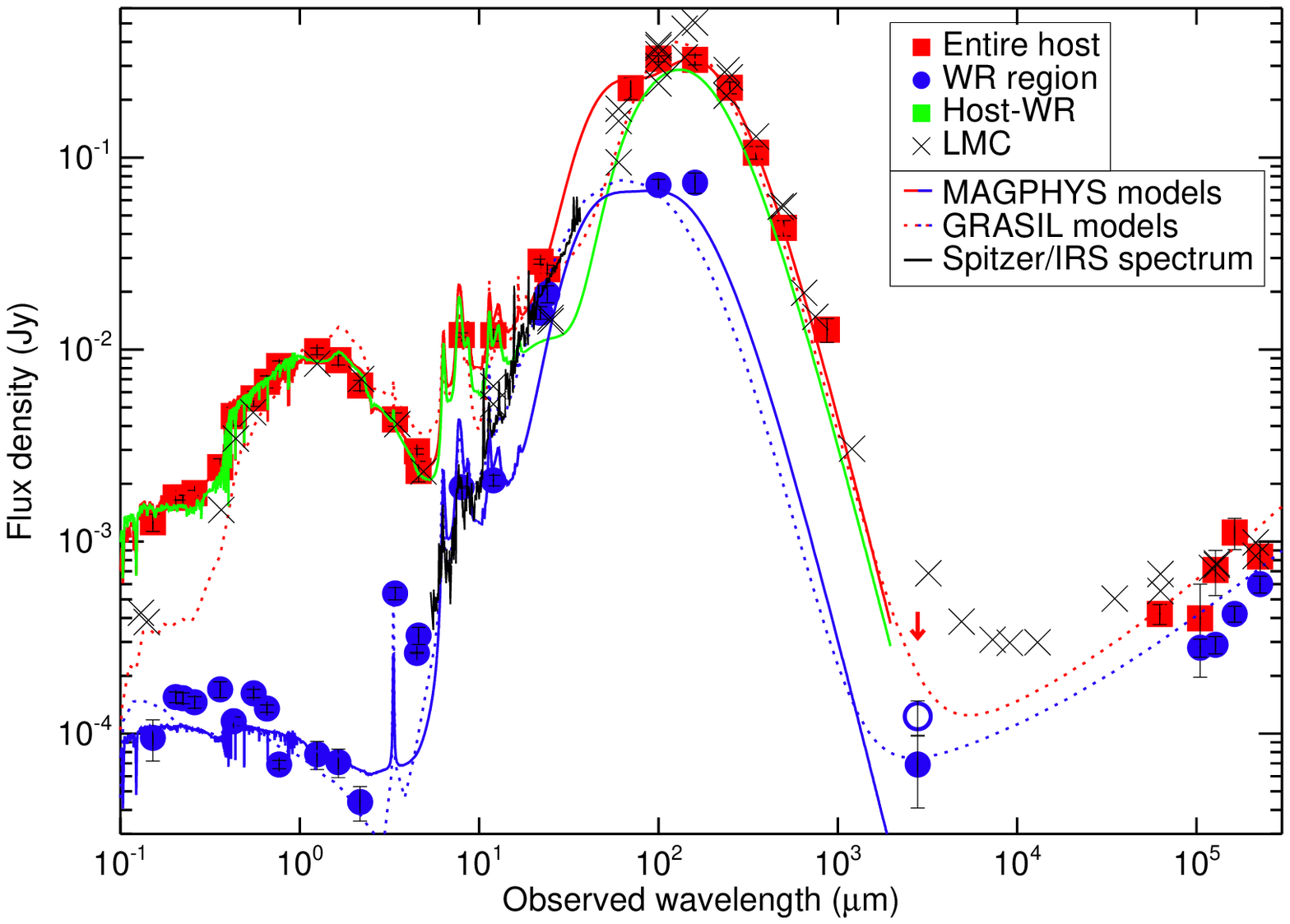}
\end{center}
\caption{Spectral energy distribution of the GRB 980425 host ({\it red}), the WR region ({\it blue}), and the host with the WR region subtracted ({\it green}). The data are shown as {\it squares} and {\it circles}, whereas the {\sc Magphys} and {\sc Grasil} models are shown as {\it solid} and {\it dotted lines}, respectively. The {\it empty blue circle} denotes the ALMA flux of the WR region without the correction for the free-free emission. { The mid-IR {\it Spitzer}/IRS spectrum of the WR region \citep{lefloch12} is shown as a {\it black line}.} { The `host$-$WR' photometry almost overlaps with the total host photometry in most cases, so it is not shown. The data for the LMC are shown as {\it black crosses} \citep{israel10,meixner13}.}
}
\label{fig:sed}
\end{figure*}

The best-fit  SEDs\footnote{The SED fits can be downloaded from 
\url{http://dark.nbi.ku.dk/research/archive}%
} 
are shown in Figure~\ref{fig:sed} and the parameters derived from the {\sc Magphys} fit are shown in Table~\ref{tab:magphysres}. { No {\sc Grasil} model was found to have sufficiently high UV emission and relatively low far-IR emission to explain the data for the entire host. However, for the WR region the {\sc Grasil} model found by \citet{michalowski09} fits our new photometry as well. The WR datapoints at $\sim0.4$--$0.6\,\micron$ are underestimated by the models, but this spectral range is dominated by strong emission lines \citep{hammer06,christensen08} not accounted for in the models.}

With the excellent wavelength coverage at the dust peak, we confirm that the WR region contributes 
$25$\%  to the infrared luminosity of the host  \citep[cf.][]{sollerman05,lefloch,lefloch12,christensen08,michalowski09}. 
{ However,  for the first time we were able to constrain its dust content and we found that it contributes only 
$\sim5$\% 
to the dust mass of the host. Its lower contribution to the dust mass compared with infrared luminosity is a consequence of its high dust temperature (both dust components in the WR region are hotter than those in the entire host, and the contribution of the hotter component is higher in the WR region, see Col.~8--11 in Table~\ref{tab:magphysres}).} 
As in \citet{michalowski09} we found that the WR region contributes very little to the stellar mass of the host.

We also performed SED modelling using the photometry of the host after subtracting the WR region contribution (green line in Fig.~\ref{fig:sed} and the last row of Table~\ref{tab:magphysres}).  Given the strong contribution of the WR region only in the mid-IR, it is unsurprising that the properties we derived from this `host$-$WR' photometry are very similar to those obtained using the original host photometry, except for much weaker contribution of hot $T_d>40$ K dust to the infrared luminosity, as the `warm' component has only $\sim35$~K for the `host$-$WR'  photometry (see Col.~10 in Table~\ref{tab:magphysres}). 
Hence, a possible contribution of WR regions in higher redshift GRBs, which cannot be separated from host galaxies, should not strongly affect the global host properties.
{ The stellar mass derived from the `host$-$WR' photometry is nominally higher than that from the `entire host' photometry  (though consistent within errors) because without the WR region the SED modelling converges to a slightly less active star-forming system with a higher mass-to-light ratio.}

Our stellar mass for the entire host 
 is $\sim2$ times lower than that derived by 
\citet{michalowski09} 	
and \citet{savaglio09} and $30$\% lower than that derived by 
\citet{castroceron10}, 
{ after correcting for the different IMF used}. These differences are, however, comparable to the expected uncertainty of stellar mass estimates \citep[see discussion in][]{michalowski12mass}. Indeed, in contrast, our mass estimates for the WR region 
 is a few times 
 higher than that of \citet{michalowski09}. 
{ The derived masses correspond to stellar mass-to-light ratios in the H- and K-bands of $\sim0.6$ and $\sim1\,\msun L_\odot^{-1}$ for the entire host, respectively and 
$\sim1.5$ and $\sim2.2\,\msun L_\odot^{-1}$ for the WR region.} 

When we fitted a modified black body to the $100$--$870\,\micron$ data for the entire host, we obtained $T_d\sim24\pm1$~K { assuming $\beta=1.5$, and $T_d\sim28\pm1$~K assuming $\beta=1$ (which provides a better fit to the $870\,\micron$ data)}. These temperatures are typical for its $L_{\rm IR}$; using the half-light radius of $r=4$ kpc we derived the infrared surface brightness $\Sigma_{\rm IR}=L_{\rm IR}/ 2 \pi r_h^2=10\,L_\odot\mbox{ pc}^{-2}$ \citep[eq.~2 of ][]{chanial07}, which  should indeed yield $T_d\sim26$~K according to Eq.~23 of \citet{chanial07}, which characterises the $T_d$-$\Sigma_{\rm IR}$ relation of local IRAS galaxies.

{ The dust mass estimates from {\sc Magphys} are based on the dust mass absorption coefficient  $\kappa(850\micron)=0.77 \mbox{ cm}^2 \mbox{ g}^{-1}$ \citep{dacunha08,dunne00}, which is approximately a factor of two higher than is assumed in the model of \citet{draine07b}. Since such lower coefficient was applied to the local galaxies observed with {\it Herschel}, used in Section~\ref{sec:nature} as a comparison sample, we recomputed the dust mass of the GRB 980425 host applying a comparable method outlined in \citet{bianchi13}, obtaining $\log(M_d/\msun)=6.57\pm 0.05$, indeed a factor of two higher than the {\sc Magphys} estimate.
}

\subsection{Flux ratios}
\label{sec:fluxr}

\begin{figure*}
\begin{center}
\includegraphics[width=\textwidth,clip]{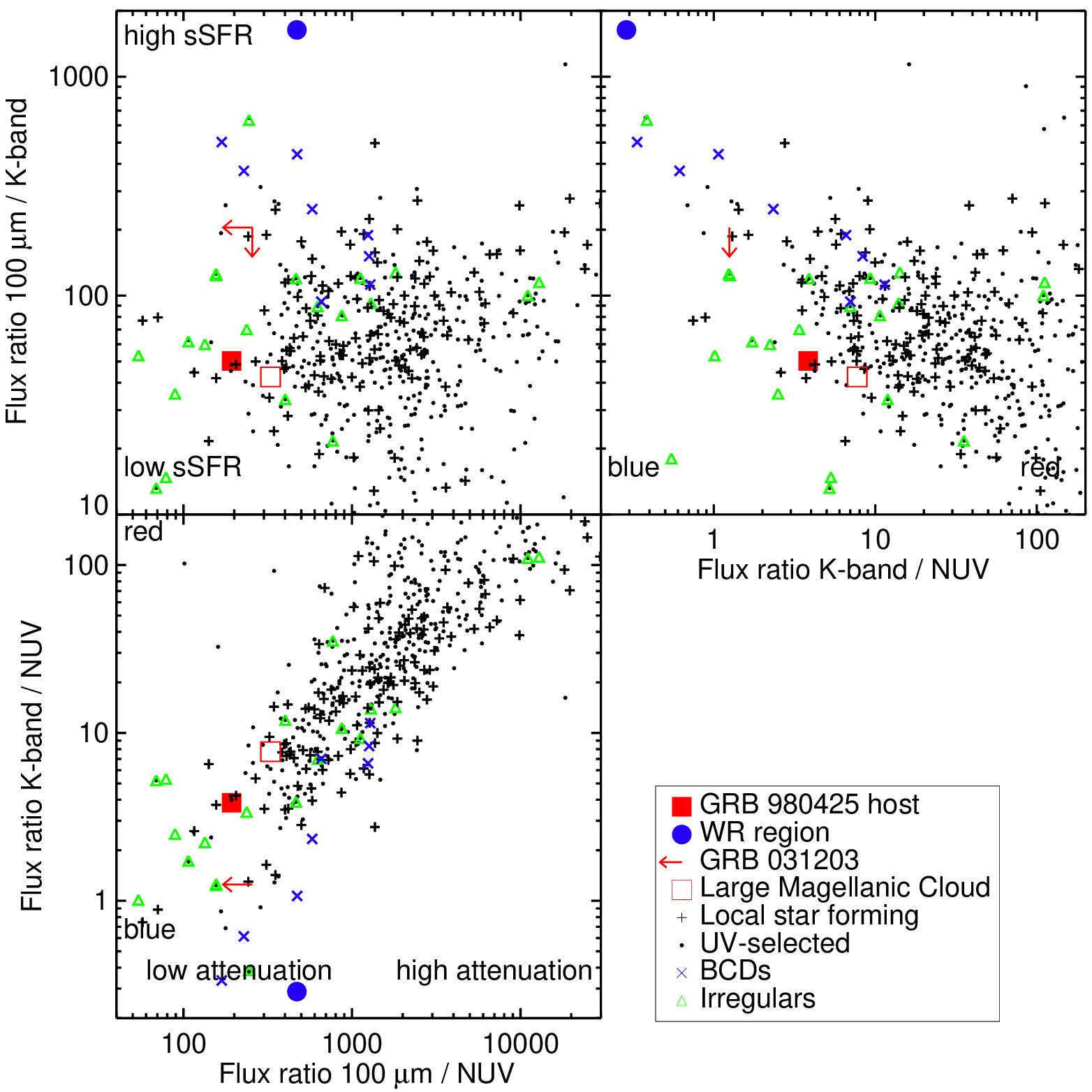}
\end{center}
\caption{Flux ratios of the GRB 980425 host ({\it red filled square}) and the WR region ({\it blue circle}) compared with GRB 031203 \citep[{\it red arrow};][]{watson11}, LMC \citep[{\it red open square};][]{israel10,meixner13},  local star-forming galaxies \citep[{\it plusses};][]{moustakas06,kingfish,dale07, dale12,skibba11}, UV-selected galaxies  ({\it dots}; \citealt{gildepaz07}; irregular galaxies are marked as green triangles), irregular galaxies ({\it triangles}) and blue compact dwarfs \citep[$\times$;][]{hunt05,hirashita08,hunter10}.
Low-$z$ GRB hosts exhibit low $100\,\micron$ / NUV flux ratios, which hints at low dust attenuation similar to that of irregular galaxies. 
}
\label{fig:fluxr}
\end{figure*}

In order to investigate the properties of the GRB 980425 host, in Fig.~\ref{fig:fluxr} we compared it with other samples of local galaxies with respect to $100\,\micron$, $K$-band, and near-ultraviolet (NUV) flux ratios, characterizing specific star formation rate (sSFR$\mbox{}\equiv\mbox{SFR}/M_*$ approximated by the $100\,\micron$ / $K$-band flux ratio), dust attenuation ($100\,\micron$ / NUV) and near-infrared / UV colour ($K$-band / NUV). We also added the GRB 031203 host at $z=0.105$ using the data from \citet{watson11}.

For comparison, we used the sample of \citet{moustakas06}, which includes UV-excess galaxies (comprising blue compact galaxies with strong emission lines, starburst nuclei, and normal massive spiral and irregular galaxies with abnormally high star formation rates), IRAS-selected galaxies, H$\alpha$- and  UV-selected galaxies within $11$ Mpc, star-forming galaxies in the Ursa Major Cluster and morphologically disturbed galaxies. Moreover, we used the sample of UV-selected galaxies of \citet{gildepaz07}, for which irregular galaxies are marked as green triangles in Fig.~\ref{fig:fluxr}. We also used the sample of irregular galaxies and blue compact dwarfs (BCDs) from \citet{hunt05}, \citet{hirashita08}, and \citet{hunter10}, for which we compiled the UV
 \citep{kinney93,dale07,gildepaz07}, $K$-band \citep{thuan83, jarrett03,hunt05,engelbracht08}, and $100\,\mu$m photometry \citep{rice88,soifer89,sanders03,lisenfeld07, hirashita08,dale09,hunter10}. { The Key Insights on Nearby Galaxies: A Far-Infrared Survey with Herschel (KINGFISH) sample \citep{kingfish,dale07, dale12,skibba11} is also used as a comparison; KINGFISH irregular galaxies are marked as green triangles in Fig.~\ref{fig:fluxr}. Dust masses for the KINGFISH galaxies are taken from \citet{bianchi13},
and stellar masses from \citet{skibba11}  are corrected to the \citet{kingfish} distance scale.} { Finally, we mark the Large Magellanic Cloud (LMC) using the compilation in \citet{israel10} and \citet{meixner13}.
}

Both GRB 980425 and 031203 hosts exhibit low $100\,\micron$ / NUV ratios (indicating low attenuation and dust content), { within the range spanned by other local galaxies, especially late types and  
irregulars}. Moreover, they have low $100\,\micron$ / $K$-band ratios (proxy for sSFR) compared with other galaxies,  given their low $K$-band / NUV ratios, { again similarly to late-type spirals}.

\begin{figure}
\begin{center}
\includegraphics[width=\linewidth,clip]{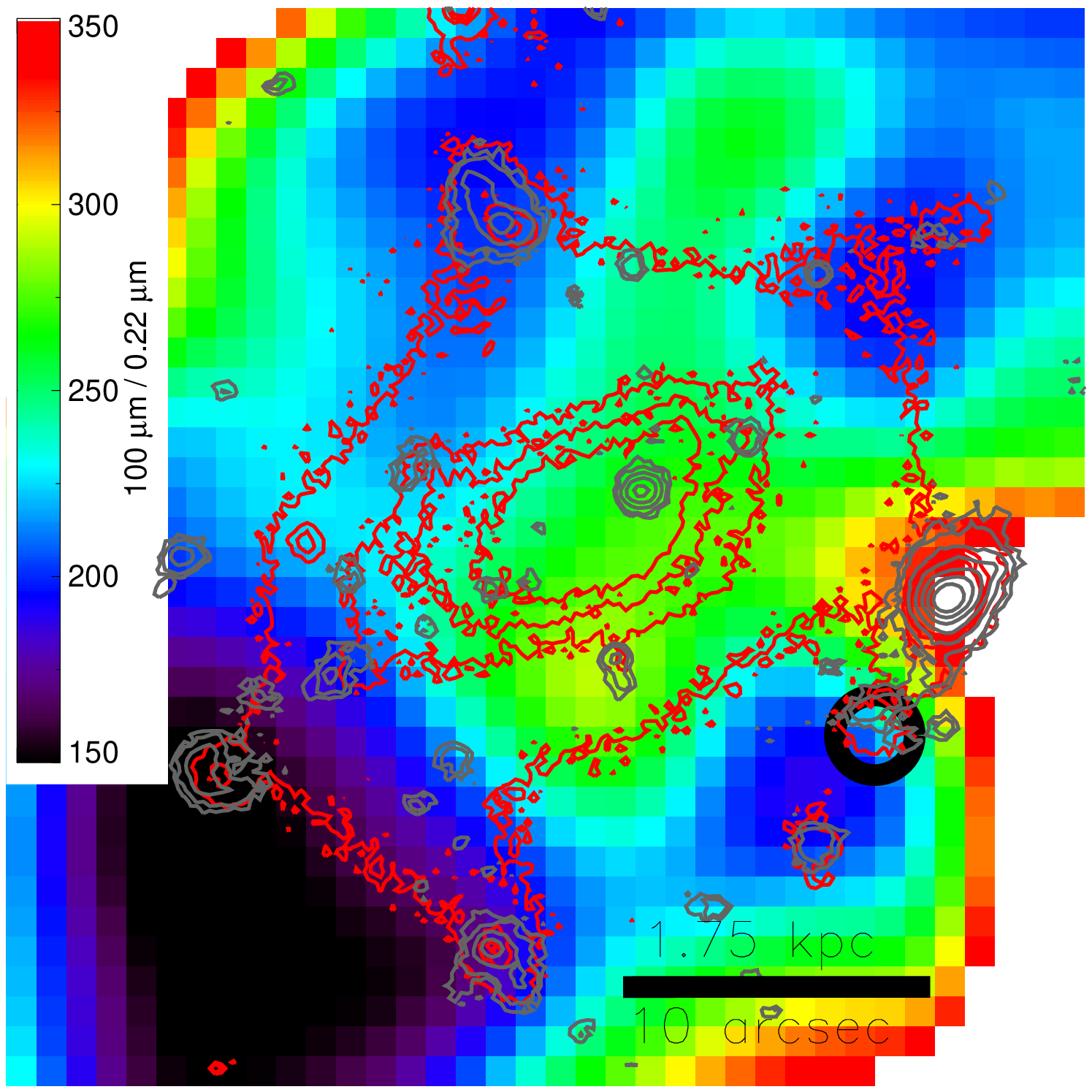}
\end{center}
\caption{$100\,\micron$ / $0.22\,\micron$ flux ratio map of the GRB 980425 host. The image is $35''\times35''$ and is constructed from  images adaptively smoothed to the same spatial scales using {\sc Asmooth} \citep{asmooth} to avoid strong colour fluctuations.  The contours represent the narrow-band H$\alpha$ map ({\it grey}, confirmed to be consistent with the IFU H$\alpha$ image from \citealt{christensen08}) and $R$-band image  \citep[{\it red};][]{sollerman05}. The {\it black circle} shows the GRB position. The WR region is to the north-west of this position. It is apparent that the low  $100\,\micron$ / NUV flux ratio (blue regions) corresponds to the spiral arms traced by  star-forming regions (peaks of red and grey countours), with the exception of the WR region exhibiting very high $100\,\micron$ flux.
}
\label{fig:colormap}
\end{figure}

In order to investigate  its low $100\,\micron$ / NUV ratio  we constructed the flux ratio map, in a similar way  to that described by \citet{boquien11,boquien12}: we converted both maps to Jy per pixel, subtracted the local sky background, corrected the NUV map for the extinction in our Galaxy using the parametrization of \citet{cardelli89} and \citet{odonnell94} with $E(\mbox{B$-$V})= 0.059$ mag, and regridded the $100\,\micron$ map to the NUV pixelscale. 
Then the $100\,\micron$ and  NUV maps were { adaptively smoothed to the same spatial scales using {\sc Asmooth} \citep{asmooth} to avoid strong colour fluctuations}, and divided. The result is shown in Fig.~\ref{fig:colormap} 
 with the narrow-band H$\alpha$ and R-band contours \citep{sollerman05} overplotted. The former was confirmed to be consistent with the integral field unit (IFU) H$\alpha$ spectroscopy from \citet{christensen08} in the part of the host covered by the IFU observations. 

From this map we found that the regions with low $100\,\micron$ / NUV ratios correspond closely to the  star-forming regions dominating the H$\alpha$ (and hence the SFR) image, with the exception of the WR region exhibiting very high $100\,\micron$ flux.

\section{Discussion}
\label{sec:discussion}

\subsection{Nature of the host: star-forming dwarf}
\label{sec:nature}

In terms of the basic properties, it has been suggested that the host of GRB 980425 is a typical dwarf galaxy \citep{sollerman05,christensen08,michalowski09}. Its SFR, stellar mass, size, and metallicity 
agree perfectly with that of the largest dwarf irregulars in the sample of \citet[][]{woo08}.  Moreover, we also found it similar to irregular galaxies in terms of far-IR/UV flux ratios (Fig.~\ref{fig:fluxr}). 
Finally, its sSFR is consistent with the high-end of samples of local galaxies   with similar stellar masses  
\citep[e.g.][]{gilbank11,gavazzi13}.
{  The metallicity of the GRB 980425 of
$12 + \log(\mbox{O}/\mbox{H})=8.6$ \citep{sollerman05}
is roughly
 consistent with
the typical value for a $10^9\,\msun$ galaxy in the mass-metallicity relations of \citet{kewley08},}
and with the mass-metallicity relation of local dwarf galaxies \citep{hunt12}. 

\begin{figure}
\begin{center}
\includegraphics[width=0.5\textwidth,clip]{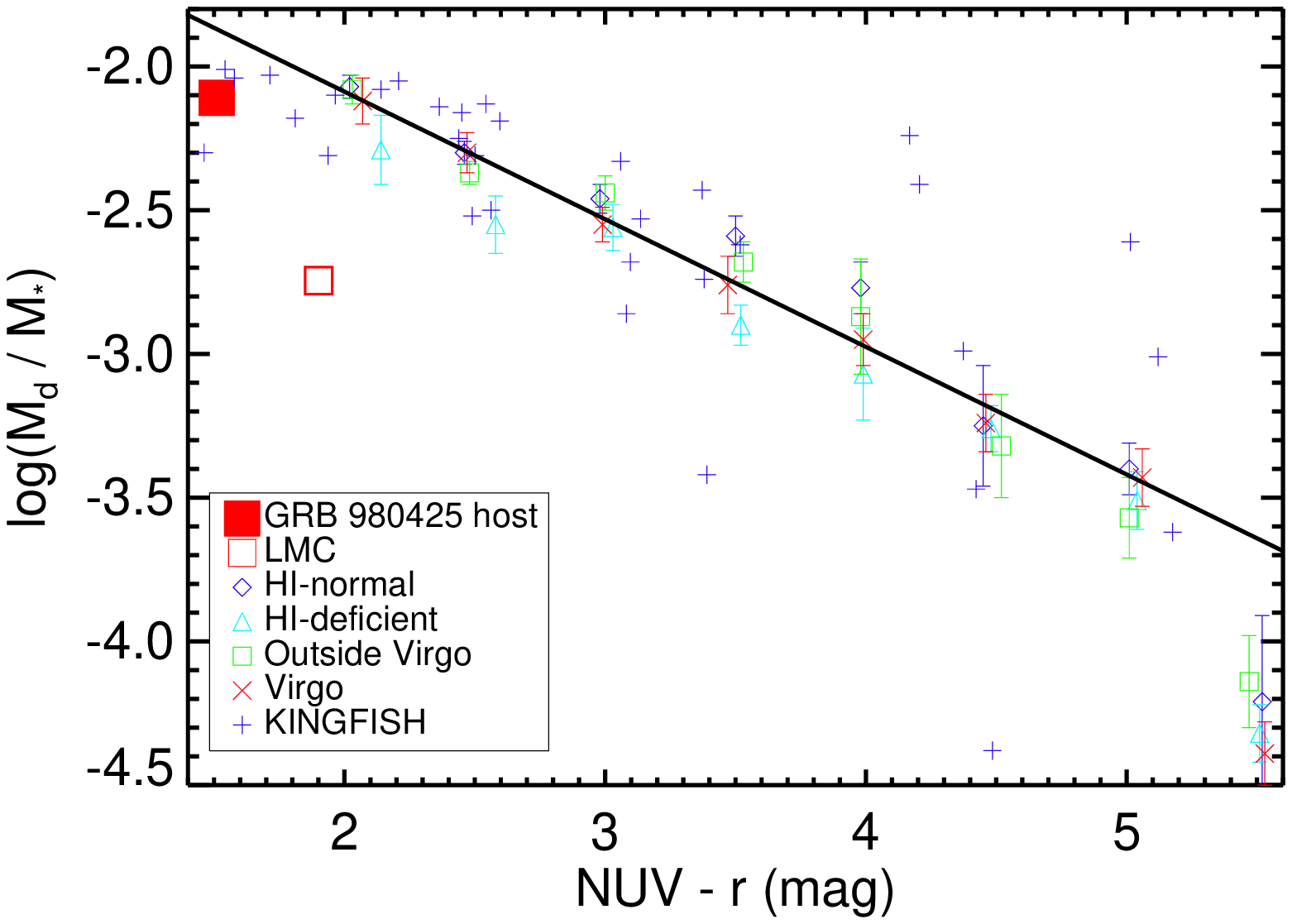}
\end{center}
\caption{Dust-to-stellar mass ratio as a function of UV-to-optical colour of the GRB 980425 host ({\it red square}), LMC 
({\it open red square}; $M_*$ from \citealt{skibba12} and $M_d$ derived applying the method of \citealt{bianchi13} to the data from \citealt{meixner13}),
KINGFISH galaxies 
\citep[{\it blue plusses};][]{kingfish},
and the averages of other local galaxies in eight colour bins  \citep[Table~1 and Fig.~4 of][]{cortese12}. The {\it solid line} represents a linear fit to the data, which results in $M_d/M_*=-1.87\pm0.07$ at the host colour of $\mbox{NUV}-r\sim1.5$, 
close to
the measured value of $M_d/M_*\sim-2.5$. 
{ The WR region is off this plot with $\mbox{NUV}-r\sim-0.15$ mag and $M_d/M_*\sim-2.1$.}
}
\label{fig:mdmsnuvr}
\end{figure}

{ We also found that the GRB\,980425 host has { relatively low dust content}, similar to local dwarf galaxies. Specifically, it exhibits a low $100\,\micron$ / NUV flux ratio (Fig.~\ref{fig:fluxr}) and a dust-to-stellar ratio $\log (M_d/M_*)\sim-2.1$ (using $M_d$ derived in Sect.~\ref{sec:prop}), consistent with that of the dwarf local galaxies in the HRS \citep[Fig.~3 of][]{cortese12}  and KINGFISH surveys. It is also consistent with the low-end of the local $M_d$--SFR relation \citep{dacunha10}. In Fig.~\ref{fig:mdmsnuvr} we also show that the GRB\,980425 host has a very blue optical/UV colour and the dust content expected for such galaxies based on the HRS and KINGFISH surveys. Moreover, the dust-obscured SFR derived from the infrared luminosity of the host (SFR$_{\rm IR}$) is only $\sim0.1\,\msunyr$ \citep[using the relation of][converted to the Chabrier IMF]{kennicutt}, which is $\sim2.5$ times lower than the total SFR from the SED modelling (Table~\ref{tab:magphysres}) and than the UV-only estimate \citep{michalowski09}. This means that most of its  star formation activity  is {\it not} dust-obscured. Finally, we demonstrate in Fig.~\ref{fig:sed} that although the SEDs of the GRB 980425 host and of the LMC are remarkably similar in the far-IR regime, the former is much brighter in the UV.

To summarise, the properties of the GRB 980425 host are consistent with those of a population of local dwarf irregular galaxies, which exhibit relatively low dust content and a high fraction of UV-visible star formation. The IRAS infrared luminosity function of \citet{sanders03} shows that $\sim90$\% of local galaxies are less luminous than the GRB 980425 host with $L_{\rm IR}\sim10^9\,\lsun$, and that $\sim15$\% of star formation activity in the local universe happens in these faint galaxies. Therefore, assuming that the GRB rate traces the SFR density, it is not surprising to find a GRB in a dwarf galaxy.
}

\subsection{Dust and star formation in the WR region: high density}
\label{sec:wr}

{ 
The WR region is the most unusual feature of the GRB 980425 host, dominating its mid-IR, far-IR, and radio emission, and is the only star-forming region detected by ALMA at $2.8$ mm (Figs.~\ref{fig:im} and \ref{fig:sed}). 
We demonstrate here that its high ISM gas density can explain these properties, as it  likely plays an important role in efficient dust production and efficient build-up of radio emission.

First, we checked whether 
its age is sufficient to form the dust that we observe in this object. Taking into account only the stars that can finish their main-sequence phase in $\sim10$ Myr (i.e. those with masses $>15\,\msun$) we calculated an average dust yield per star in order to explain the dust mass we measured for the WR region \citep[as in][]{michalowski10smg4,michalowski10qso}. Assuming a Chabrier IMF, its stellar mass corresponds to $43\,000$ of these stars. Hence, each of them should have produced $\sim1.4\,\msun$ of dust. This is close to the highest dust yields predicted theoretically by \citet[][Fig.~12]{nozawa03} and \citet[][Fig.~6]{gall11c}, 
{ and inferred observationally \citep{dunne03,dunne09casA,morgan03b,gomez09,green04, ercolano07,meikle07,lee09,sibthorpe09,barlow10}}.
Given the possibility of rapid ($\sim10$ Myr) dust grain growth in the ISM \citep{draine79,draine90,draine09,hirashita00,zhukovska08,michalowski10qso}, which would lower the required dust yield per star, it is not unreasonable to assume that the dust present in the WR region could have been formed during a short ($\sim10$ Myr) recent period of star formation, even if it had contained no pre-existing dust { formed by AGB stars}. 
{ Moreover, these massive stars would alone produce enough metals necessary for this dust production (not counting metals produced by older AGB stars). Indeed a SN produces $\lesssim1\,M_\odot$ of heavy elements \citep{todini01,nozawa03,bianchi07,cherchneff09}, which is close to what is required to explain the dust mass in the WR region.}

The timescale of dust grain growth is inversely proportional to the gas density \citep[eq.~8 of][]{draine09} and is only $\sim2$ Myr at a density of $150\,\mbox{cm}^{-3}$. Hence, the unusually high far-IR luminosity (Fig.~\ref{fig:im}) and dust mass of the WR region could be explained if it is the densest among the star-forming regions in the GRB 980425 host.

This is consistent with the finding of \citet{christensen08} who reported slightly higher electron density for this region compared with other star-forming regions. We re-calculated the densities  of all regions from the [\ion{S}{II}] line ratios \citep{christensen08} using the calibration of \citet{odell13} based on the model of \citet{osterbrock06}: $\log n_e =4.705-1.9875  [\ion{S}{II}\,6716] /  [\ion{S}{II}\,6731]$. We obtained $n_e=150$ and $100\,\mbox{cm}^{-3}$ for the WR region and the SN site, respectively \citep[consistently with][]{christensen08} and the mean for all regions of $130\pm15\,\mbox{cm}^{-3}$. As shown in Fig.~\ref{fig:im}, five out of seven other regions with high densities $n_e> 145\,\mbox{cm}^{-3}$ are located  in the northern spiral arms, which is the third most  prominent  $100\,\micron$ feature after the WR region and the central bar \citep[the northernmost $100\,\micron$ blob is not covered by the IFU data of][]{christensen08}. This supports the scenario in which high density of the ISM is responsible for more efficient dust production. 

These other star-forming regions with high $n_e$ are all younger than the WR region (Fig.~\ref{fig:ne}). This provides an explanation for why their $100\,\micron$ emission is not as prominent as that of the WR region, as there was even less time to form dust. On the other hand, other regions (some older than the WR region) are not dense enough to accumulate dust efficiently. { However, the  regions older than the WR region (green crosses in Figs.~\ref{fig:im} and \ref{fig:ne}) are all concentrated in the bar, which shows pronounced $100\,\micron$ emission.}

The high density of the WR region would also explain why it is the only region firmly detected at the radio continuum (Fig.~\ref{fig:im}). 
{ According to \citet[][]{hirashitahunt06} the radio fluence of supernova remnants depends on gas density, naturally explaining the radio brightness of the WR region if it is dense.  On the other hand, the other regions denser than the WR region are also younger (Fig.~\ref{fig:ne}), so their synchrotron radiation might not have had time to build up yet. }
The WR region indeed exhibits a steep radio spectrum (Fig.~\ref{fig:sed} and Sect.~\ref{sec:irradio}), hinting at the synchrotron nature of the emission.

{
Finally, the ALMA detection of the WR region at 2.8\,mm is also consistent with its high density. 
Free-free emission coefficients are proportional to the square of the density  \citep{condon} making the low density regions fainter
than the WR.
The dust-only emission is a $\sim 2.5\sigma$ detection, so without the free-free emission (45\% of the WR 2.8\,mm emission)
the WR region would probably not have been detected.
}

The WR region appears to be similar to the 30 Doradus (Tarantula Nebula), which also dominates the H$\alpha$ \citep{gaustad01}, far-IR \citep{meixner06}, and radio \citep{dickel05} emission of the  LMC,  as shown in Fig.~\ref{fig:lmc}, 
but it contributes very little to total stellar mass. The star-forming region 30 Doradus is very dense  \citep[with a density close to the value we measured for the WR region; e.g.][]{meixner06,pellegrini10,pellegrini11,kawada11,pineda12}, which provides additional support for our hypothesis that the extreme properties of the WR region are due to its high density.

If high density environments are also found close  to the positions of other GRBs, then the ISM density should also be considered, along with metallicity, an important factor influencing whether a given stellar population can produce a GRB. This is especially true if the  \citet{hammer06} `runaway' hypothesis, that the GRB progenitor was born in the WR region and expelled from it, turns out to be correct. The properties of the WR region we discussed here (high SFR, high density, numerous massive stars, low age) are consistent with this hypothesis, but  high-resolution observations of other GRB hosts are needed to test whether GRB progenitors are usually born in dense star clusters.

\begin{figure}
\begin{center}
\includegraphics[width=0.5\textwidth,clip]{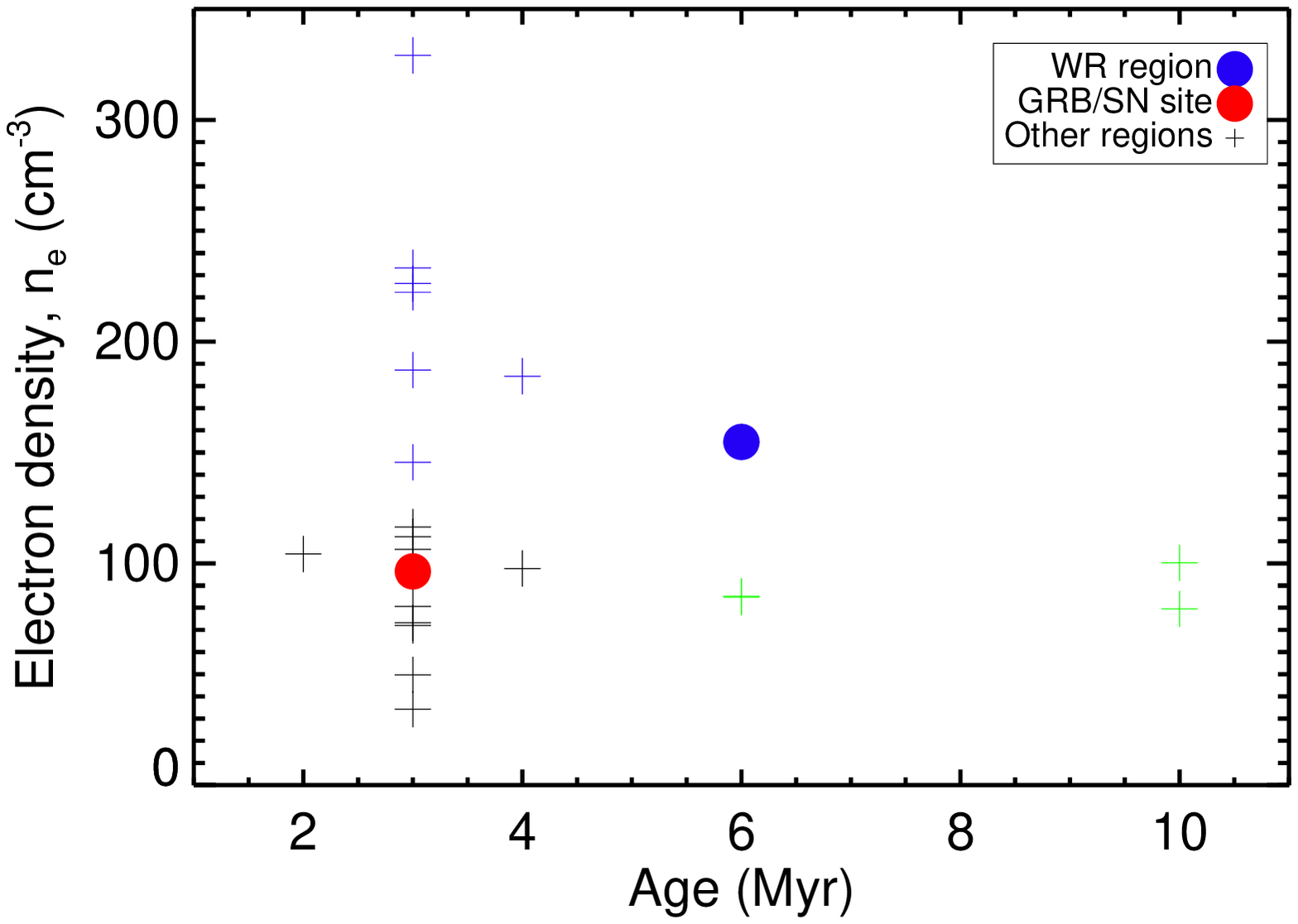}
\end{center}
\caption{Electron density of the star-forming regions in the GRB 980425 host (derived using the [\ion{S}{II}] flux ratio from \citealt{christensen08} using the calibration of \citealt{odell13} and \citealt{osterbrock06}) as a function of their ages from the SED modelling \citep{christensen08}. The WR region is both old and dense compared with other regions, which may have helped it to accumulate dust more efficiently. {\it Blue} and {\it green crosses} mark the regions with $n_e>145\,\mbox{cm}^{-3}$ and $\mbox{age}\ge6$ Myr, respectively (see also Fig.~\ref{fig:im}).
}
\label{fig:ne}
\end{figure}

\begin{figure}
\begin{center}
\includegraphics[viewport=220 10 380 830,clip,angle=-90,width=0.5\textwidth,]{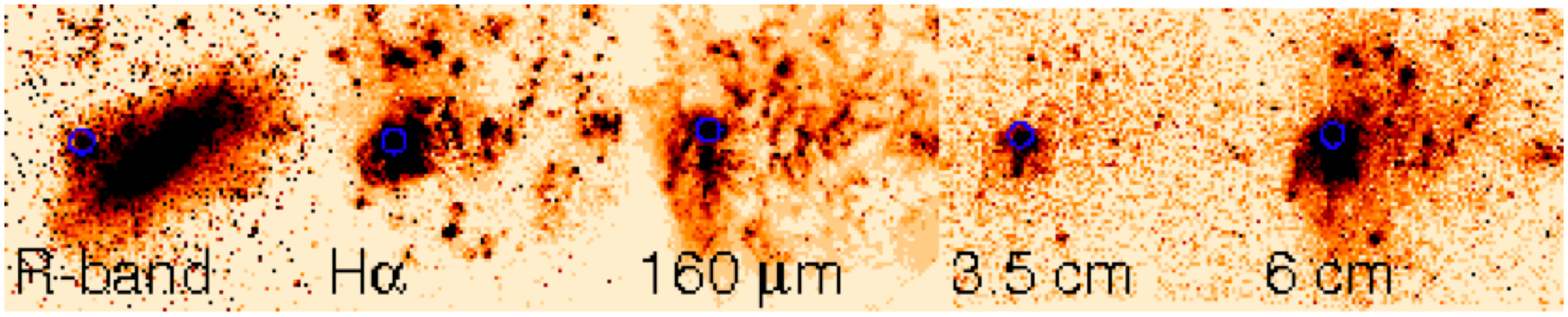}
\end{center}
\caption{Mosaic of the images of  the Large Magellanic Cloud \citep{gaustad01,meixner06,dickel05}. North is up and east is to the left. Each panel is  $6^\circ\times6^\circ$  ($4\mbox{ kpc}\times4\mbox{ kpc}$).  The {\it blue  circle}  show the position of the star-forming region 30 Doradus (Tarantula Nebula). The LMC with 30 Doradus dominating at H$\alpha$, far-IR and radio is similar to the GRB 980425 host with the WR region (compare with Fig.~\ref{fig:im}).
}
\label{fig:lmc}
\end{figure}

\subsection{Submillimetre excess and gas-to-dust ratio}
\label{sec:excess}

Many dwarf low-metallicity galaxies have been shown to exhibit a {\submm} excess, i.e. an enhancement in the {\submm} emission above the extrapolation from far-IR data \citep{lisenfeld02, galliano03, galliano05, galliano11, bendo06, galametz09,galametz11, bot10, gordon10,israel10, planck11,planck11b,dale12, delooze12}. Apart from the cold dust emission, this excess has also been attributed to different grain properties { (e.g.~shallower spectral slope $\beta$)}, spinning dust, and magnetic nanograins (e.g.~\citealt{reach95}, \citealt{draine07b}, \citealt{dale12}, \citealt{hermelo13} and references therein).

The GRB 980425 host exhibits a $160\,\micron$ / $870\,\micron$ flux ratio of $\sim25$, { lower than} the lowest values found by \citet[][]{galametz11} for local dwarfs, indicating a very strong {\submm} excess. 
This is not because of the  synchrotron contribution to the {\submm} flux, which should be minimal given the flux of $0.42$ mJy at  $6$ cm  and 
 a negative radio slope \citep[see Sect.~\ref{sec:irradio} and][]{michalowski09},
nor is it due to contamination by free-free emission because, as we found in Sect. \ref{sec:sed}, it contributes at most 2\% to the total
870\,\micron\ flux.

Given the far-IR wavelength coverage, we are in a position to investigate the {\submm} excess in the GRB 980425 host. { We present two hypotheses to explain this excess: very cold dust and unusual dust properties. Under the assumption of normal dust properties,} based on the SED fit (Fig.~\ref{fig:sed}) we predict an $870\,\micron$ flux of  $\sim7.5$ mJy. This is about half the measured value of $\sim13$ mJy (Table~\ref{tab:data}). If this excess is interpreted as a contribution of very cold ($10$ K) dust, not accounted for in our SED modelling, then it transforms into an additional $2\times10^6\,\msun$ of dust (with $\beta=1.5$), doubling the dust content in the GRB 980425 host.

Without this cold dust, its molecular gas-to-dust ratio is $M_{{\rm H}_2}/M_d<185$ using the H$_2$ gas mass upper limit from \citet{hatsukade07}. This is consistent with that of the Milky Way  \citep[$\sim100$--$400$;][]{sodroski97,draine07} and other spirals  \citep[$\sim100$--$1000$;][]{devereux90,stevens05}. However, if the {\submm} excess is indeed due to cold dust, then the limit for the GRB 980425 host lowers to $M_{{\rm H}_2}/M_d<82$. This indicates much higher dust content, consistent with that of high-redshift {\submm} galaxies \citep[$\sim50$;][]{kovacs06,michalowski10smg4}, of the nuclear regions of local (ultra)luminous IR galaxies \citep[$120\pm28$;][]{wilson08} and of local, far-IR-selected galaxies \citep[$\sim50$;][]{seaquist04}. 
This can be confirmed by a detection of a CO line of the GRB 980425 host to constrain the molecular gas content, and by investigation of the {\submm} excess at other wavelengths { and at higher resolution}, allowing a more precise {study of its origin.  This is now possible with the current capabilities of ALMA.}

The additional cold dust is not the only possible explanation of the {\submm} excess in the GRB 980425 host. As indicated in Section~\ref{sec:prop}, we are able to fit all far-IR and {\submm} data with a modified black-body curve with a shallow spectral slope of $\beta=1$ without the need to invoke cold dust (see Fig.~13 of \citealt{bendo06} for a similar result for another galaxy). In this scenario, a very low $160\,\micron$ / $870\,\micron$ flux ratio of the GRB 980425 host would be a result of unusual dust properties favouring emission at longer wavelengths.
Alternatively, such a shallow slope could indicate dust heating by an unusually low-intensity radiation
field which results in a wider range of dust temperatures along the line of sight (Hunt et al., in prep.).

Our low-resolution $870\,\micron$ data do not allow the study of the spatial distribution of the {\submm} excess. However, ALMA data provide a hint that the WR region may be responsible for the excess. 
Nevertheless, because almost half of the observed $2.8$\,mm flux is free-free emission, sensitive
high-resolution {\submm} observations, probing cold dust emission, are necessary to confirm this.

\subsection{Infrared-radio correlation}
\label{sec:irradio}

Based on the  extrapolation from the $24\,\micron$ data, \citet{michalowski09} suggested that the radio emission of the GRB 980425 host is much lower than expected from the infrared-radio correlation \citep{condon}. However, with the far-IR wavelength coverage provided by our {\it Herschel} data we revised the infrared luminosity estimate to be a factor of $\sim2.5$ lower. In order to quantify the offset of the host from the infrared-radio correlation, we calculated the parameter $q=\log(L_{\rm IR}[L_\odot] / 3.75\times10^{12}/L_{\nu\, 1.4\, {\rm GHz}} [L_\odot \mbox{Hz}^{-1}])$, where $L_{\nu\, 1.4\, {\rm GHz}}$ is the rest-frame $1.4$ GHz luminosity density computed from { the power-law fit to our radio data and the $4.8$ GHz data of \citet{michalowski09}. We find a spectral index of $\alpha=-0.62\pm0.15$ \citep[consistent with a typical value of $-0.75$ for other star-forming galaxies][]{condon,clemens08,dunne09,ibar10} and a rest-frame $1.4$ GHz flux of $0.91\pm0.17$ mJy, which corresponds to a $L_{\nu\,\rm{rest}\, 1.4\, {\rm GHz}}=1.5\times10^{27}$ erg s$^{-1}$ Hz$^{-1}$. This translates to $q=2.8\pm0.5$, fully consistent with the average value for local star-forming galaxies of $2.64$ \citep[with a scatter of $0.26$;][]{bell03}. 

In a similar way, for the WR region we derived a steeper $\alpha=-1.08\pm0.18$,  $F_{\nu\,\rm{rest}\, 1.4\, {\rm GHz}}=0.56\pm0.08$, $L_{\nu\,\rm{rest}\, 1.4\, {\rm GHz}}=8.9\times10^{26}$ erg s$^{-1}$ Hz$^{-1}$, and $q=2.48\pm0.37$. This $q$ value is 
also
consistent with that of local galaxies within the errors. A slightly lower $q$ value may indicate that the radio emission in the WR region is building up faster than the dust content.

The mid-infrared and radio (close to the rest-frame $1.4$ GHz) data presented by \citet{watson11} for  the GRB 031203 host ($L_{\rm IR}\sim2\times10^{10}\,L_\odot$, $L_{\nu\, 1.4\, {\rm GHz}}\sim10^{29}$ erg s$^{-1}$ Hz$^{-1}$) also implies a low $q=2.25$. 
However, the data of \citet{watson11} extends only up to $30\,\micron$, so increased wavelength coverage is required to confirm this result.
}

\section{Conclusions}
\label{sec:conclusion}

{
Using  high-resolution observations from {\it Herschel}, APEX, ALMA, and ATCA we investigated the properties of the ISM in the GRB 980425 host,
and we found that it is characterised by relatively low dust content and a high fraction of UV-visible star formation, similar to other dwarf galaxies. These galaxies are abundant in the local universe, so it is not surprising to find a GRB in one of them, assuming the correspondence between the GRB rate and star formation.

The star-forming region displaying the Wolf-Rayet signatures in the spectrum (WR region), located $800$ pc from the GRB position, contributes substantially to the host emission  at the far-infrared, millimetre, and radio wavelengths and we propose this to be a consequence of its high gas density. If dense environments are also found close to the positions of other GRBs, then the ISM density should also be considered, along with metallicity, as an important factor influencing whether a given stellar population can produce a GRB, especially if it turns out that GRB progenitors in general (and that of GRB 980425 in particular) are born in dense star clusters}. The brightness of the WR region also indicates that it may be responsible for the submillimetre excess we detected for the host.

\begin{acknowledgements}

We thank Joanna Baradziej for help with improving this paper; Tom Muxlow, Eelco van Kampen, and Robert Braun for help with the ALMA and ATCA observations; Lise Christensen for kindly providing the IFU maps from \citet{christensen08}; our anonymous referee, Vincent H\'{e}nault-Brunet, and  Fran\c{c}oise Combes for useful comments; and  Elisabete da Cunha for providing updated filter profiles for {\sc Magphys}.

MJM and GG are postdoctoral researchers of the FWO-Vlaanderen (Belgium).
MJM acknowledges the support of the Science and Technology Facilities Council. 
LKH and SB are supported by the INAF PRIN 2012 grant.
The Dark Cosmology Centre is funded by the Danish National Research Foundation.
TM and DB acknowledge the support of the Australian Research Council through grant DP110102034.
AdUP acknowledges support from the European Commission (FP7-PEOPLE-2012-CIG 322307) and from the Spanish project AYA2012-39362-C02-02.

PACS has been developed by a consortium of institutes led by MPE (Germany) and including UVIE (Austria); KU Leuven, CSL, IMEC (Belgium); CEA, LAM (France); MPIA (Germany); INAF-IFSI/OAA/OAP/OAT, LENS, SISSA (Italy); IAC (Spain). This development has been supported by the funding agencies BMVIT (Austria), ESA-PRODEX (Belgium), CEA/CNES (France), DLR (Germany), ASI/INAF (Italy), and CICYT/MCYT (Spain). 
SPIRE has been developed by a consortium of institutes led by Cardiff University (UK) and including Univ. Lethbridge (Canada); NAOC (China); CEA, LAM (France); IFSI, Univ. Padua (Italy); IAC (Spain); Stockholm Observatory (Sweden); Imperial College London, RAL, UCL-MSSL, UKATC, Univ. Sussex (UK); and Caltech, JPL, NHSC, Univ. Colorado (USA). This development has been supported by national funding agencies: CSA (Canada); NAOC (China); CEA, CNES, CNRS (France); ASI (Italy); MCINN (Spain); SNSB (Sweden); STFC (UK); and NASA (USA). 
This paper makes use of the following ALMA data: ADS/JAO.ALMA\#2011.0.00046.S. ALMA is a partnership of ESO (representing its member states), NSF (USA) and NINS (Japan), together with NRC (Canada) and NSC and ASIAA (Taiwan), in cooperation with the Republic of Chile. The Joint ALMA Observatory is operated by ESO, AUI/NRAO and NAOJ. 
This publication is based on data acquired with the Atacama Pathfinder Experiment (APEX). APEX is a collaboration between the Max-Planck-Institut fur Radioastronomie, the European Southern Observatory, and the Onsala Space Observatory.  
The Australia Telescope is funded by the Commonwealth of Australia for operation as a National Facility managed by CSIRO. 
This research has made use of data obtained from the High Energy Astrophysics Science Archive Research Center (HEASARC), provided by NASA's Goddard Space Flight Center. 
This publication makes use of data products from the Wide-field Infrared Survey Explorer, which is a joint project of the University of California, Los Angeles, and the Jet Propulsion Laboratory/California Institute of Technology, funded by the National Aeronautics and Space Administration. 
This research has made use of 
the GHostS database (\urltt{http://www.grbhosts.org}), which is partly funded by Spitzer/NASA grant RSA Agreement No. 1287913; 
the NASA/IPAC Extragalactic Database (NED) which is operated by the Jet Propulsion Laboratory, California Institute of Technology, under contract with the National Aeronautics and Space Administration;
SAOImage DS9, developed by Smithsonian Astrophysical Observatory \citep{ds9};
and the NASA's Astrophysics Data System Bibliographic Services.

\end{acknowledgements}



\end{document}